\documentclass[fleqn,usenatbib]{mnras}

\usepackage{newtxtext,newtxmath}

\usepackage[T1]{fontenc}

\DeclareRobustCommand{\VAN}[3]{#2}
\let\VANthebibliography\thebibliography
\def\thebibliography{\DeclareRobustCommand{\VAN}[3]{##3}\VANthebibliography}

\usepackage{graphicx}	
\usepackage{amsmath}	

\def\ud{\mathrm{d}}

\newcommand{\orcid}[1]{\href{https://orcid.org/#1}{\,\includegraphics[width=8px]{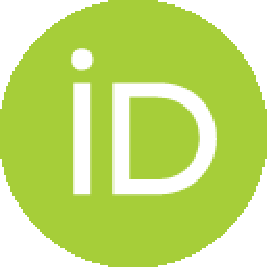}}}

\title[Physical vs phantom dark energy after DESI]{Physical vs phantom dark energy after DESI:\\ thawing quintessence in a curved background}

\author[B. R. Dinda et al.]{
Bikash R. Dinda\orcid{0000-0001-5432-667X},$^{1,2}$\thanks{E-mail: bikashrdinda@gmail.com}
Roy Maartens\orcid{0000-0001-9050-5894}$^{1,2}$
\\
$^{1}$Department of Physics \& Astronomy, University of the Western Cape, Cape Town 7535, South Africa\\
$^{2}$National Institute for Theoretical \& Computational Science, Cape Town 7535, South Africa\\
}

\date{Accepted XXX. Received YYY; in original form ZZZ}

\pubyear{2015}

\begin{document}
\label{firstpage}
\pagerange{\pageref{firstpage}--\pageref{lastpage}}
\maketitle

\begin{abstract}
Recent data from DESI, in combination with other data, provide moderate evidence of dynamical dark energy, $w\neq-1$. In the $w_0, w_a$ parametrization of $w$, there is a preference for a phantom crossing, $w<-1$, at redshift $z\sim0.5$. In general relativity, the phantom equation of state is unphysical. Thus it is important to check whether phantom crossing is present in other physically self-consistent models of dark energy that have equivalent evidence to the $w_0, w_a$ parametrization. We find that thawing quintessence with nonzero cosmic curvature can fit the recent data as well as $w_0, w_a$ in a flat background, based on both parametric and realistic scalar field evolutions. Although the realistic model does not allow $w<-1$, the parametrizations do allow it. However even if we allow $w<-1$ the data do not enforce phantom crossing.  
Thus, the phantom crossing is an artifact of a parametrization that is not based on a physical model.
\end{abstract}




The Dark Energy Spectroscopic Instrument (DESI) Data Releases DR1 and DR2  \citep{DESI:2024mwx,DESI:2024aqx,DESI:2024kob,DESI:2025zgx,Lodha:2025qbg} have delivered state-of-the-art precision on baryon acoustic oscillation (BAO) measurements. This has facilitated stringent constraints on the dark energy equation of state, $w$ \citep{Wolf:2025jed,Park:2024pew,Park:2024vrw,Park:2025azv,Colgain:2025nzf,RoyChoudhury:2024wri,choudhury2025,Chakraborty:2025syu,Dinda:2025svh,Dinda:2024ktd,Dinda:2024kjf,Gialamas:2024lyw}. Using the parametrization
\begin{equation}
w(z)=w_0+w_a\,\frac{z}{1+z}\,,
\label{eq:cpl}
\end{equation}
both DESI DR1 and DR2 BAO data favour a phantom crossing, from a phantom value $w<-1$ at earlier times, $ z\gtrsim 0.5$, to $w>-1$ for $ z\lesssim0.5$. Phantom behavior is unphysical in general relativity since energy conservation implies the growth of dark energy density with expansion, $\dot\rho_{\rm de}>0$. 

It is implied by similar parametrizations which are not based on physical models of $w$ (e.g., see Fig. 4 in \cite{Lodha:2025qbg}). It is true that models of interacting dark energy in general relativity \citep{Valiviita:2008iv} can have an {\em effective} $w_{\rm eff}<-1$ while the intrinsic $w$ is $\ge -1$. Similarly, modifications to general relativity \citep{Clifton:2011jh} can mimic phantom behaviour. However before abandoning simple physical models in general relativity, we need to check whether phantom behaviour is in fact required by the data.

For a physically motivated model, DESI data suggest that we consider dark energy models which behave close to a cosmological constant $w=-1$ at early times, with an increase to $w>-1$ at late times (e.g., see Figs. 11 and 12 in \citep{Lodha:2025qbg}). Physically consistent examples are provided by
the `thawing' class of quintessence models \citep{Linder:2007wa,Linder:2015zxa}.

\begin{figure}
\centering
\includegraphics[height=135pt,width=0.46\textwidth]{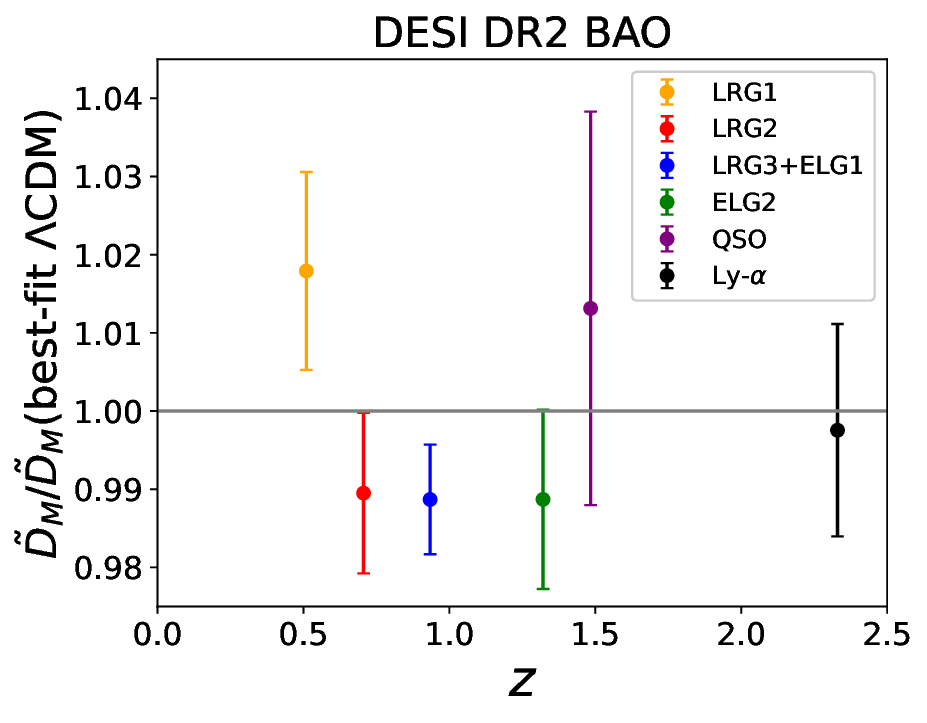}
\includegraphics[height=140pt,width=0.46\textwidth]{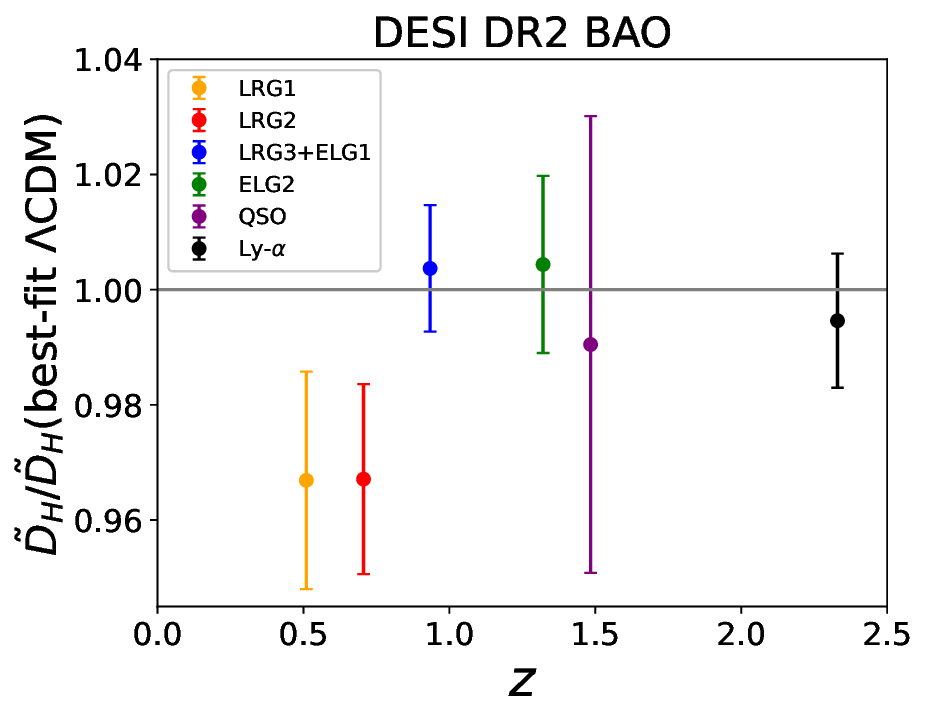}
\caption{
\label{fig:flat}
Comparison of $\tilde{D}_M=D_M/r_d$ (upper panel) and $\tilde{D}_H=D_H/r_d$ (lower panel) to the best-fit values of  flat $\Lambda$CDM, obtained from CMB+DESI DR2 BAO.
}
\end{figure}

\begin{figure}
\centering
\includegraphics[height=140pt,width=0.46\textwidth]{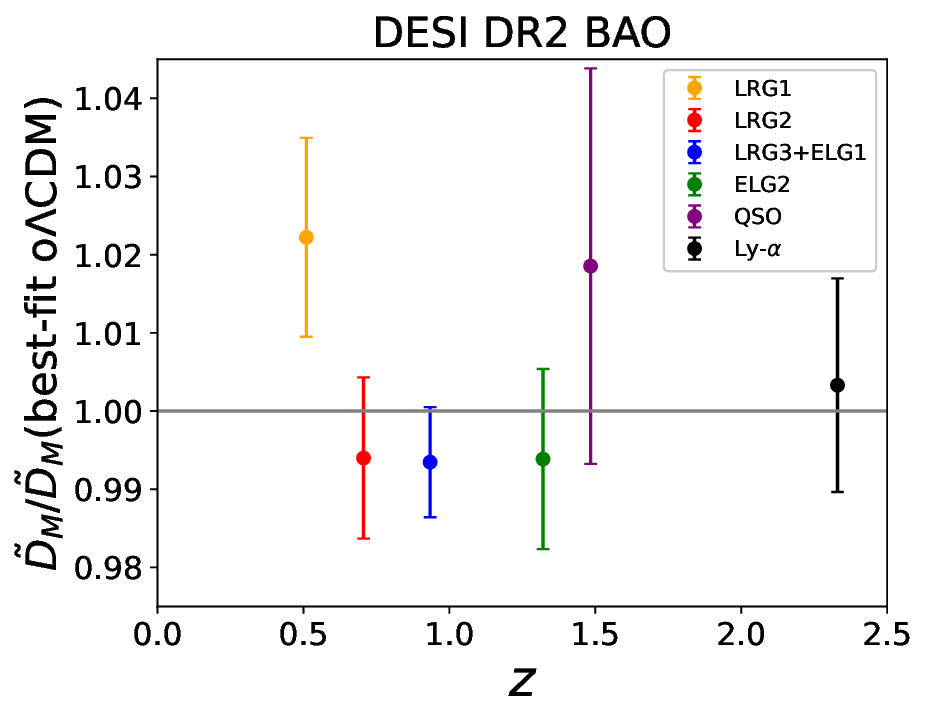}
\includegraphics[height=140pt,width=0.46\textwidth]{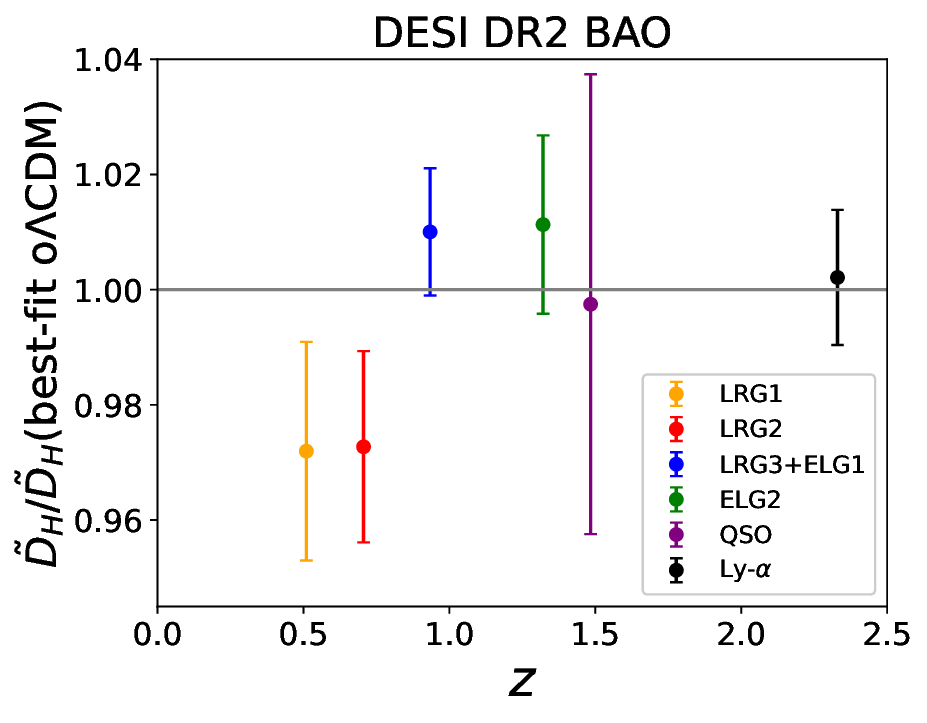}
\caption{
\label{fig:curved}
As in \autoref{fig:flat} but for best-fit o$\Lambda$CDM values.
}
\end{figure}

To investigate the robustness of this phantom crossing, we start by showing the DESI DR2 BAO data for $\tilde{D}_M=D_M/r_d$ and $\tilde{D}_H=D_H/r_d$ (where $D_M$ and $D_H$ are comoving and Hubble distances and $r_d$ is the comoving sound horizon at baryon drag). We also show the best fit $\Lambda$CDM in the flat (\autoref{fig:flat}) and negatively curved o$\Lambda$CDM (\autoref{fig:curved}) cases, using also the CMB distance prior, as in DESI DR2 \citep{DESI:2025zgx} (see their Appendix A). From \autoref{fig:flat} and \autoref{fig:curved} it is clear that at earlier times ($z\gtrsim 1.5$) the deviations from both $\Lambda$CDM and o$\Lambda$CDM model are well within 1$\sigma$. At intermediate times ($0.9\gtrsim z\gtrsim 1.5$) the deviations are $\sim1\sigma$, and at late times $ z\lesssim0.9$ the deviations are $\lesssim 2\sigma$. Note that these results are similar to the lower panels of Fig. 6 in \citep{DESI:2025zgx}. These plots hint that there is no evidence for deviation from  $\Lambda$CDM (or o$\Lambda$CDM)  at higher redshifts ($z\gtrsim 1.5$). At intermediate redshifts ($0.9\gtrsim z\gtrsim 1.5$) there is no significant deviation either. This is evidence that the phantom crossing is a model-dependent artifact of the $w_0w_a$CDM model. Evidently, the $w_0w_a$CDM or similar parametrizations are too simplistic to correctly capture the behavior at lower and higher redshifts simultaneously and they may worsen the $H_0$ tension \citep{Colgain:2025nzf}. Here we focus on the implications of DESI data for dynamical dark energy, without considering the possible problems introduced by systematics (see e.g. \cite{Sapone:2024ltl,Colgain:2025nzf,Colgain:2025fct,Colgain:2024mtg}).

\autoref{fig:flat} and \autoref{fig:curved} also reveal another interesting feature. In flat  $\Lambda$CDM, $\tilde{D}_H=\tilde{D}'_M$ so that the deviations in $\tilde{D}_M$ and $\tilde{D}_H$ should be similar in the flat case. However, DESI DR2 data shows that the deviations are different in $\tilde{D}_M$ and $\tilde{D}_H$ at intermediate redshifts. The DESI DR2 main paper \citep{DESI:2025zgx} finds the constraint  $10^3 \Omega_{\rm K0}=2.3\pm1.1$ on cosmic curvature (their Table~V). Of course, inferences about curvature may not hold for dynamical dark energy, since in that case, $w$ is partially degenerate with $\Omega_{\rm K0}$. Cosmic curvature 
directly affects  distances by altering the geometry of photon paths \citep{Liu:2020pfa,Wei:2019uss,Dinda:2023kvg}. Because it evolves as $(1+z)^2$,  curvature has its strongest relative effect in the narrow redshift window between matter domination and dark energy domination, when its influence is seen only through its contribution to the energy budget.

At this stage, our task is half complete, because the intermediate to higher redshift behavior can be well modeled by the o$\Lambda$CDM model, but definitely not the lower redshift results. The lower redshift data is what allows $w_0w_a$CDM or similar parametrizations to show a hint of deviations from $\Lambda$CDM and of phantom crossing even in the presence of nonzero cosmic curvature. The o$\Lambda$CDM model cannot properly capture the behavior over the entire redshift range either. Consequently, our next step is to find a physically consistent model which can capture the behavior for the whole redshift range.

We consider quintessence dark energy models, which have self-consistent physical properties, including $w\ge -1$ and a speed of sound $c_s=1$ \citep{Caldwell:1997ii,Caldwell:2005tm,Tsujikawa:2013fta}. Quintessence models cover a huge range of behavior, but we focus on the thawing class (TQ) which naturally produces $w>-1$ at low redshift, as suggested by the predictions in $w_0w_a$CDM and similar parametrizations; see Fig.~4 in \cite{Lodha:2025qbg}. We also include cosmic curvature (oTQ). For generality, we use two  parametrizations of $w$ which approximate  thawing quintessence models \citep{Carvalho:2006fy,Linder:2015zxa,Linder:2007wa,Shlivko:2025fgv,Abreu:2025zng,Akthar:2024tua}:
\begin{eqnarray}
w(z) &=& -1 + \frac{1+w_0}{(1+z)^3} \,,
\label{eq:oTQ1} \\
w(z) &=& -1 + \frac{3^{2/3}(1+w_0) \big(z^3 +3z^2 + 3z+3\big)^{-2/3}}{1+z}  .~~~
\label{eq:oTQ2}
\end{eqnarray}
We identify these 1-parameter $w$ models as oTQ1 and oTQ2.
The Hubble rate is given by
\begin{eqnarray}
\frac{H^2}{H_0^2} &=& \Omega_{\rm r0}(1+z)^4+\Omega_{\rm m0}(1+z)^3+\Omega_{\rm K0}(1+z)^2 \nonumber\\
&&{} + \Omega_{\rm q0} \exp \left[ 3 \int_0^z \frac{1+w(\tilde{z})}{1+\tilde{z}} d\tilde{z} \right] ,
\label{eq:Esqr} 
\end{eqnarray}
where
$\Omega_{\rm r0}+\Omega_{\rm m0}+\Omega_{\rm K0}+\Omega_{\rm q0} = 1$.

\begin{figure}
\centering
\includegraphics[height=170pt,width=0.47\textwidth]{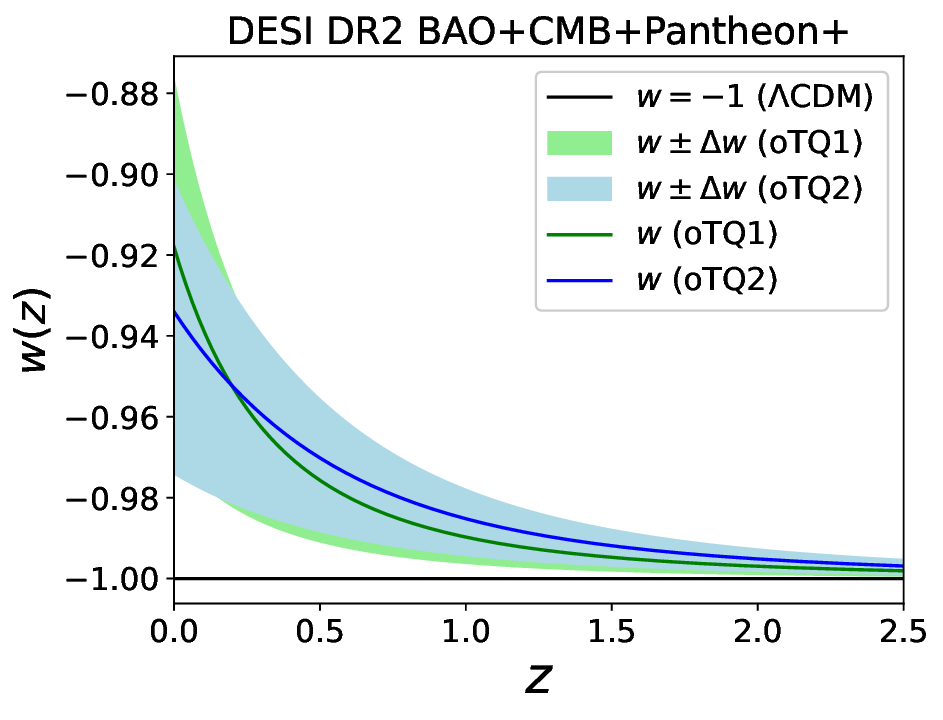}
\caption{
\label{fig:eos}
Thawing quintessence models oTQ1,2. Constraints on $w_0$ are from DESI DR2 BAO\,+\,CMB\,+\,Pantheon+.
}
\end{figure}

\autoref{fig:eos} shows the mean $w(z)$ for oTQ1,2 as given by \autoref{eq:oTQ1} and \autoref{eq:oTQ2}, with 1$\sigma$ shaded regions. Constraints are from DESI DR2 BAO+CMB+Pantheon+.  The full (realistic) quintessence models (described later), with $w\equiv w_\phi=p_\phi/\rho_\phi$, do not allow $w<-1$, but this is not automatically enforced in the $w(z)$ parametrizations oTQ1,2 above. However, even if we impose a prior $w_0<-1$ in the data analysis, it does not alter the results in \autoref{fig:eos}. This shows that there is no evidence of phantom crossing in these models.  

In order to make a stronger statement, we should show that the oTQ models are at the same (or better) evidence level as $w_0w_a$CDM. To this end, we compare the best-fit log-likelihood values of these models:
\begin{align} \label{lnl}
\ln L \mbox{\,(oTQ1,2)} - \ln L (w_0w_a {\rm CDM}) = (0.47,~0.37)\,.
\end{align}
Since the number of parameters is the same in all three models ($w_0, \Omega_{\rm K0}$ in oTQ1,2), the best-fit log-likelihood comparison is enough to show the model comparison. We see that oTQ1 and oTQ2 are slightly preferred over $w_0w_a$CDM. Although this is not significant, it is enough to show that the phantom crossing is not inevitable or even preferred.

Note that in general, models of quintessence worsen the $H_0$ tension \citep{Vagnozzi:2019ezj,Banerjee:2020xcn,DiValentino:2021izs,Dinda:2021ffa,Lee:2022cyh}, but we checked the constraints on $H_0$ and found no significant differences between these three models. Thus, this tension is not improved or worsened by the thawing quintessence models with nonzero cosmic curvature.

A dark energy parametrization is a simple and useful way to do data analysis in cosmology -- but preferably we need to show that a physically motivated model of dark energy, in this case a realistic thawing quintessence, produces similar results \citep{DESI:2024kob,Akrami:2025zlb,Borghetto:2025jrk,Shajib:2025tpd,Wolf:2025jlc,Dubey:2025cdk,Tada:2024znt,Ramadan:2024kmn,Wolf:2024eph,Berbig:2024aee,Goldstein:2022okd,Payeur:2024dnq,Velten:2014nra,Steinhardt:2025znn}. We consider a class of realistic thawing quintessence models in the presence of cosmic curvature. In the presence of matter, radiation, and curvature, a general quintessence field obeys the evolution equations \citep{Clemson:2008ua,Dinda:2017swh,Bhattacharya:2024hep,Andriot:2024jsh}:
\begin{eqnarray}
\frac{\ud w}{\ud N} &=& (w-1) \Big(3w +3 -\lambda  \sqrt{3(w+1)  \Omega _{\rm q }}\,\Big) , \nonumber\\
\frac{\ud\Omega_{\rm q}}{\ud N} &=& \Omega_{\rm q } \big[3 w \Omega _{\rm q }-3w -\Omega _{\rm K}+\Omega _{\rm r}\big] , \nonumber\\
\frac{\ud\lambda}{\ud N} &=& -\sqrt{3} (\Gamma -1) \lambda ^2 \sqrt{(w+1) \Omega _{\rm q }} \,, \nonumber\\
\frac{\ud\Omega_{\rm r}}{\ud N} &=& \Omega _{\rm r} \big[3w \Omega _{\rm q }-\Omega _{\rm K}+\Omega _{\rm r}-1\big] , \nonumber\\
\frac{\ud\Omega_{\rm K}}{\ud N} &=& \Omega _{\rm K} \big[3 w \Omega _{\rm q }-\Omega _{\rm K}+\Omega _{\rm r}+1\big] ,
\label{eq:quint} 
\end{eqnarray}
where  $N=\ln a$, the slope of the quintessence potential is $\lambda=-M_{\rm pl} {\partial_{\phi}V}/{V}$ and $\Gamma={V \partial^2_{\phi}V}/ {(\partial_{\phi}V)^2}$.
For non-constant $\Gamma$ and which cannot be explicitly expressed in terms of $\lambda$, one needs further differential equations that depend on the nature of the potential. We restrict our analysis to constant $\Gamma$, which includes exponential ($\Gamma=1$) and mononomial ($\Gamma=1+1/n$ for $V(\phi)=V_0 \phi^{-n}$) potentials.

For thawing quintessence, initial values should be $0<w_{\rm i}+1 \ll 1$ and $\lambda_{\rm i}>0$ \citep{Scherrer:2007pu,Clemson:2008ua,Chiba:2012cb,Tsujikawa:2013fta,Dinda:2016ibo}. We use $w_{\rm i}=10^{-4}-1$ and $\Gamma=5$ ($n=1/4$). We name this model oTQR (curved thawing quintessence realistic). 

\begin{figure}
\centering
\includegraphics[height=135pt,width=0.46\textwidth]{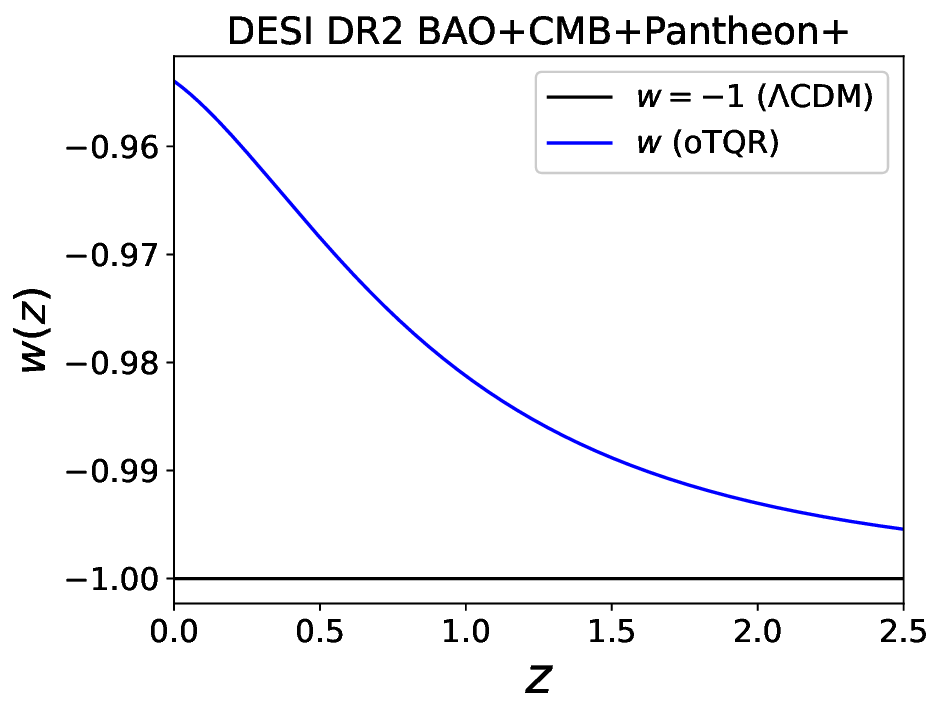} \\
\includegraphics[height=135pt,width=0.46\textwidth]{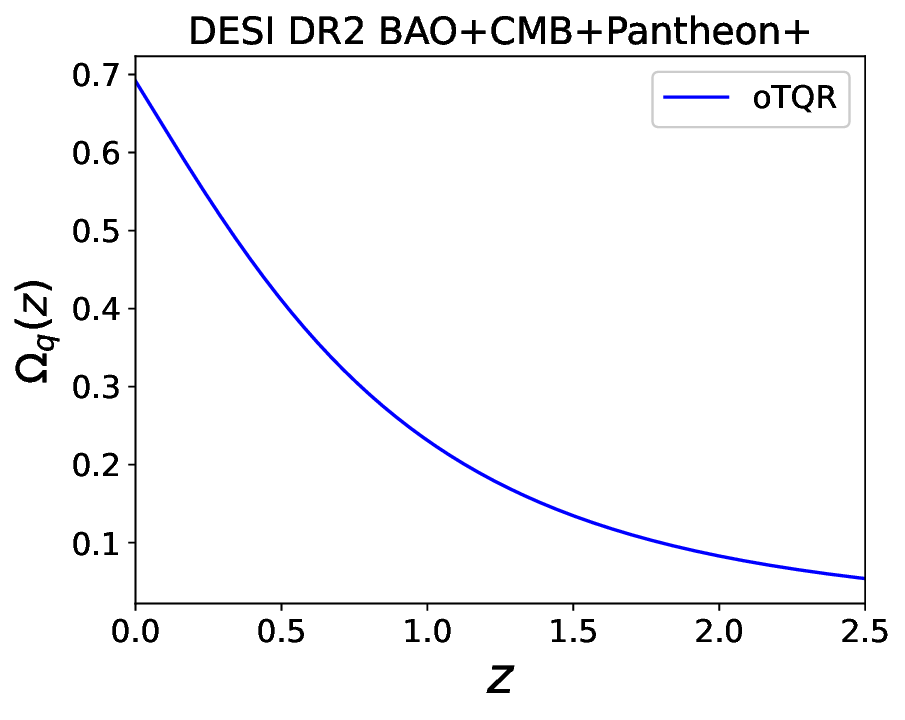}
\caption{
\label{fig:eos2}
$w(z)$ (upper panel) and $\Omega_{\rm q}(z)$ (lower panel) for a realistic thawing quintessence model (oTQR) with best-fit values obtained from DESI DR2 BAO+CMB+Pantheon+.
}
\end{figure}

\begin{figure*}
\centering
\includegraphics[height=400pt,width=0.95\textwidth]{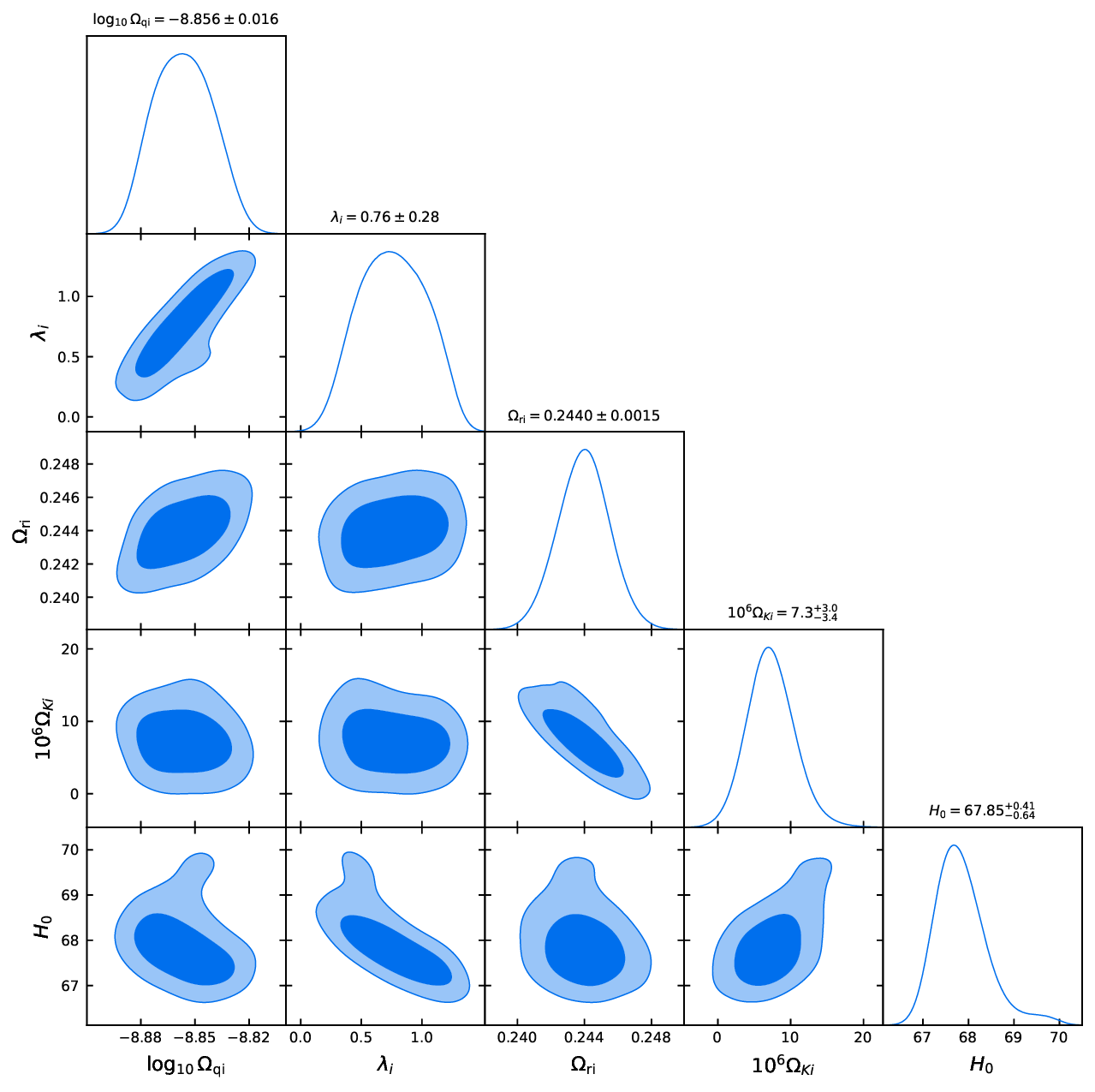}
\caption{
\label{fig:realquint}
Constraints on the oTQR model parameters for DESI DR2 BAO+CMB+Pantheon+ data.
}
\end{figure*}

Using the best-fit cosmological density parameters from DESI DR2 BAO+CMB+Pantheon+ data, we can solve the system in \autoref{eq:quint}. The resulting $w(z)$ and $\Omega_{\rm q}(z)$ are shown in \autoref{fig:eos2}.          
 \autoref{fig:realquint} shows constraints on the model parameters through triangle plots. 

In order to find proper constraints on realistic quintessence parameters, we need to have data-informed priors on these parameters \citep{Lodha:2025qbg}. However,  this is true only for the energy density parameters $\Omega_{\rm q i}$, $\Omega_{\rm ri}$, and $\Omega_{\rm Ki}$, not for the other parameters. This is due to the cosmic coincidence problem, present in all realistic physical models \citep{Zlatev:1998tr,Sahni:1999gb,Velten:2014nra}. If we try to mimic $w_0w_a$CDM by a physically motivated model, we will encounter the same coincidence problem. Thus this fact cannot be a negative point for model comparison.

It is also argued in \cite{Lodha:2025qbg} that for realistic quintessence models to fit the data well, they must have a rapid increase of $w$ for $z \lesssim 0.3$. However, the upper panel of \autoref{fig:eos2} shows that this statement is not applicable. Furthermore, it is argued that there might be features of a sharp increase in the energy density of dark energy -- but the lower panel of \autoref{fig:eos2} shows no sharp increase.

Finally, we compare the best-fit log-likelihood of the realistic open thawing quintessence model oTQR to that of $w_0w_a$CDM:
\begin{align} \label{lnl2}
\ln L \mbox{\,(oTQR)} - \ln L (w_0w_a {\rm CDM}) = -0.03\,.
\end{align}
This shows that the evidence for curved realistic quintessence is effectively equivalent to that of flat $w_0w_a$CDM.
We conclude that curved thawing quintessence is equivalently favored compared to $w_0w_a$CDM in light of DESI DR2 BAO data (in combination with other relevant observations). The evidence for phantom crossing is model-dependent.

\section*{Acknowledgements}

We are supported by the South African Radio Astronomy Observatory and the National Research Foundation (Grant No. 75415).

\section*{Data Availability}

This study did not generate new data. All analyzed data are publicly available and have been properly cited.



\bibliographystyle{mnras}
\bibliography{refs_mnras} 








\bsp	
\label{lastpage}
\end{document}